\documentstyle[prb,aps,epsfig,multicol]{revtex}

\draft

\long\def\singlecol#1{
\twocolumn[\hsize\textwidth\columnwidth\hsize\csname @twocolumnfalse\endcsname
              #1]}

\ifpreprintsty 
\long\def\singlecol#1{#1}
\fi 

\def\ve#1{ {\mathbf #1}}
\def\beq{\begin{equation}}
\def\eeq{\end{equation}}

\def\eqref#1{(\ref{#1}) }

\begin{document}
\title{Stripes and superconducting pairing in
the $t$-$J$ model with Coulomb interactions}

\author{ E. Arrigoni$^1$, 
A.~P. Harju$^1$, W. Hanke$^{1,2}$,
 B. Brendel$^1$,  and S.~A. Kivelson$^3$}
\address{$^1$Institut f\"ur Theoretische Physik, Universit\"at
  W\"urzburg, D-97074 W\"urzburg, Germany}
\address{$^2$Institute for Theoretical Physics, Univ. of California, Santa Barbara, California 93106}
\address{$^3$
Department of Physics, University of California, Los Angeles,
California 90095
}

\singlecol{
\date{\today}
\maketitle

\begin{abstract}
 We study the competition between long- and short-range interactions among
 charge carriers in strongly-correlated electronic systems employing a
 new method which combines the density-matrix renormalization-group
 technique with a self-consistent treatment of the long-range interactions.
 We apply the method to an extended t-J model which exhibits 
 ``stripe'' order. The Coulomb interactions, while not destroying
 stripes, induce large transverse stripe fluctuations with 
associated charge delocalization.
This leads to a substantial Coulomb-repulsion-induced
 {\it enhancement} of long-range superconducting pair-field correlations.
\end{abstract}

\pacs{PACS numbers:
 71.10.-w, 
 74.20.-z, 
 71.45.Lr 
}

}

\section{Introduction}
\label{intr}
Much of the interesting physics of the high-temperature
superconductors, in particular that related to the
``mechanism'' of high temperature superconductivity, is
moderately local, involving physics on the length scale of
the superconducting coherence length, $\xi_0$.  Since $\xi_0$ is typically
a few lattice constants, this would seem to indicate that numerical
solutions of model problems on clusters with as few as 50-100 sites should
be able to provide considerable insight concerning these problems, even
though results in this range are manifestly sensitive to the choice of boundary
conditions and other finite-size effects.  Such studies
can also serve as important tests of the {\it predictions} of
analytic theories.

Studies of $t$-$J$ systems have, indeed, provided strong evidence of a
universal and robust d-wave character of local pairing
correlations\cite{poil},  and of a strong clustering tendency of holes,
which might either lead to ``stripe'' ({\it i.e.} unidirectional charge
density order) formation\cite{wh.sc}, or phase
separation\cite{phasesep}.  Both of these features were in fact anticipated
by analytic theories\cite{zaanen,lo.em.94}.  However, these studies
for the $t-J$ and parent Hubbard models
have failed to find compelling evidence of the strong superconducting
correlations\cite{as.ha.94} needed to understand high temperature
superconductivity.  Moreover, many features of the results, especially with
regards to stripes, appear very sensitive to small changes in the model, {\it
e.g.} the shape and size of the cluster\cite{Hell_vs_White}, 
whether or not a small
second neighbor hopping $t'$ is included or not\cite{wh.sc},
 etc:

All these
calculations omit the 
long-range part of the Coulomb
interaction because it is difficult to treat using any of the standard
numerical methods.
However, since in the high-T$_c$ cuprates, where interactions are
generally conceded to be strong, the (inter-band)
screening is semiconductor- and not metallic-like, so 
there is no a priori justification for neglecting 
the long-range part.
In addition, in the case of stripes, charge inhomogeneities or phase
separation, longer-range Coulomb interactions are clearly 
important, a point which has previously been addressed with various
mean-field approximations~\cite{lo.em.94,se.ca.98,ha.wa.00u}

In this paper,
  we present a new computational method for studying the ground-state
  properties of electrons with strong short- and
  long-range interactions on fairly large finite systems. 
  The method, which may be termed a
  ``density-functional DMRG'', uses numerically very accurate DMRG
 methods to treat the short-range part of the 
  interactions.
 The long-range piece is
taken into account
 within the
Hartree approximation,  
which becomes exact in the long-distance limit, and, as we show below
and in Sec.~\ref{appa}, 
turns out to work well already for short distances.
 An important point to 
note is that, through the self-consistency requirement, the Hartree 
potential accounts for screening effects in a similar spirit as 
density-functional theory.

We have studied $N\times 4$ $t$-$J$ ladders and
cylinders, with hole densities per site  $n_h=1/9$, $1/8$, and
$1/7$, and $N=27, 18, 16$, and $14$.  
 Our principal findings, as summarized in the figures,
are:  1)  Charge stripe formation is robust to the inclusion of Coulomb
interactions of reasonable strength, although 
the associated charge density modulations
have their magnitude somewhat reduced.
However, ``spin stripe''
correlations ({\it i.e.} spin 
density modulations which suffer a $\pi$ phase shift across the
charge stripe resulting in a spatial period twice that of the
charge  modulations) are prominent on cylinders, but
very weak  on ladders;  they are
slightly enhanced by the Coulomb interactions.  2)  The inclusion of Coulomb interactions
strongly {\it enhances} the superconducting pair-field correlations at the
longest distances accessible in these calculations.  This is our most
striking result. 3)  We present
 evidence that stripe formation does
not suppress
{\it local} superconducting pairing.
On the other hand,
rigid stripe ordering competes with long-range phase
ordering~\cite{emery97,wh.sc}.
  The enhanced
superconducting correlations at long-distances produced by Coulomb
interactions are, thus, tentatively associated with the enhanced pair tunneling
between stripes produced by the increased stripe fluctuations, in agreement with
phenomenological 
arguments\cite{kivelson98}.  

This paper is organised as follows. In Sec.~\ref{mode} we introduce
our model and the combined DMRG-Hartree technique to deal with it. In
Sec.~\ref{resu} we present our results for the charge and spin
densities as well as for the pairing susceptibility. Sec.~\ref{conc}
discusses our conclusions. For the sake of completeness, a
discussion of the validity of the Hartree approximation is given in Sec.~\ref{appa}.
 
\section{Model and technique}
\label{mode}  
The Hamiltonian of the $t$-$J$ model 
with nearest-neighbor hopping $t$ and exchange interaction $J$ 
plus long-range Coulomb
interactions, which operates in the subspace of no doubly occupied sites, is
given by
\begin{eqnarray}
\label{ham}
H&=&-t \sum_{\langle \ve r,\ve r^{\prime} \rangle \sigma}
(c^\dagger_{\ve r\sigma} c_{\ve r^{\prime}\sigma} +  {\mathrm h.c.}) \\
 + & &J \sum_{\langle \ve r,\ve r^{\prime} \rangle}({\bf S}_{\ve r} \cdot {\bf
S}_{\ve r^{\prime}}-\frac{n_{\ve r}n_{\ve r^{\prime}}}{4}) + \sum_{\ve r}
{V}_{\mathrm Coul}  ({ \ve r})n_{\ve r} 
\ ,\nonumber
\end{eqnarray}
where $\langle \ve r,\ve r^{\prime}\rangle$ are nearest-neighbor sites, $\sigma$
is the spin index, $c^\dagger$ is the electron creation operator, ${\bf S}$ is
the spin operator, and $n_{ \ve r}= \sum_{\sigma} c^\dagger_{ \ve r \sigma} c_{ \ve r
  \sigma}$.
 The long-range part of the interaction~\cite{lrcoul} 
\begin{equation}
  {V}_{\mathrm Coul} ({ \ve r}) =  
V_0  \; 
  \sum_{ \ve r^{\prime} \ne  \ve r} \frac{n( \ve r') -
    \bar{n}}{|{ \ve r}-{ \ve r'}|} \label{coul} \ .
\end{equation}
is
treated in the self-consistent Hartree approximation, whereby the
density operator 
$n_{ \ve r}$
is replaced with its ground-state expectation value $n({ \ve  r})
\equiv <n_{\ve r}>$ and
 $\bar n$ is the uniform
 positive background charge-density.
 The Coulomb prefactor $V_0$ is given by
$V_0={\mathrm e}^2/[4\pi\varepsilon_0\,    \varepsilon \, a]$,
 where $a$ is the lattice
constant, and $\ve r$ is the coordinate of a lattice site in units of
$a$.
  This long-range potential is screened by a background dielectric
constant $\varepsilon$, given both by electronic interband and phonon
contributions, 
which we take to be
$\varepsilon=8.5$, in reasonable accordance with cluster
calculations for Coulomb matrix elements\cite{hy.st.92,emery97} and
with Quantum-Monte-Carlo simulations~\cite{sc.gr.98u}.
Thus, for $a\approx 3.4$~{\AA}, the Coulomb prefactor
in \eqref{coul} is $V_0 \approx \, t $, being of the same order of
magnitude as
the kinetic energy.

Since ${V}_{\mathrm Coul}$ and the density $n({\ve r})$ depend on
each other, one needs a self-consistent solution of 
\eqref{ham}-\eqref{coul}. Our theoretical
treatment is a combination of the DMRG
technique~\cite{dmrg}, which accurately accounts for short-range
interactions, and a Hartree treatment of long-range interactions.
 In the spirit of 
``density-functional theory'', we iteratively solve
\eqref{ham} and \eqref{coul} as follows: in the first step we set
${V}_{\mathrm Coul}=0$ and perform a DMRG calculation. This gives rise
to a density profile $\{ n({\ve r}) \}$ and, via
\eqref{coul}, to a new potential ${V}_{\mathrm Coul}$, which
enters the next step DMRG calculation as an additional on-site
potential. 
 This procedure should be repeated iteratively up to convergence~\cite{newton}.
The fact that the Hartree approximation 
is appropriate 
for the long-range Coulomb
part 
is discussed in detail in Sec.~\ref{appa}.

Within a given loop, the goal of the DMRG simulation is to
find iteratively an eigenstate of the Hamiltonian~\eqref{ham},
 using only a
fraction of all the possible states of the system~\cite{dmrg}.
  We have typically
kept around 800-1000 states in the last iterations of the
calculation, which
results in a maximum discarded weight of the order $10^{-4}$.
 We use
systems with open (and cylindrical) boundary conditions 
chosen not to frustrate the domain
walls.

\section{Results}
\label{resu}
In the end, we are able to compute ground-state energies for given
quantum numbers 
and ground-state correlation
functions.  Because we do not have excited-state data, we are unable to
compute dynamic susceptibilities.
 In the figures, we present representative results for various
ground-state correlation functions.  So as to make our principal findings
clear, we have averaged  these quantities  over the
transverse coordinate of the ladder. 
 Thus, we define the average
hole and spin density   as a function of position along the ladder 
$0<x\le N$ as
$\rho(x) \equiv \sum_{y=1}^4 [1-n(x,y)]/4$, and
$S_z(x) \equiv \sum_{y=1}^4 (-1)^{x+y}<S^z(x,y)>/4$.
Note that translational (in the $x$ direction) and spin rotational symmetry  
are explicitly broken in these calculations by the
ladder ends, themselves, and (following White and Scalapino (WS)~\cite{wh.sc}) 
by an applied
staggered Zeeman field of magnitude $h=0.1 t$ on
the ladder ends~\cite{zeeman,wh.sc}

To probe superconductivity, we have computed the ground-state
pair-field correlation function
\begin{equation}
D(x) = \left\langle \Delta \left(\frac{N}{2} + \frac{x}{2}\right) \ \Delta^\dagger \left(\frac{N}{2} - \frac{x}{2}\right)  \right\rangle \ .
\end{equation}
Here $\Delta^\dagger (x)$ 
creates a $d_{x^2-y^2}$
-like pair around the  $(x,2)$
site~\cite{pair}.
We have explored the dependence of our results on parameters to
some extent, but in all the figures have adopted a conventional value 
$J/t=0.35$. 

\begin{figure}[htb]
\psfig{width=0.4\textwidth,file=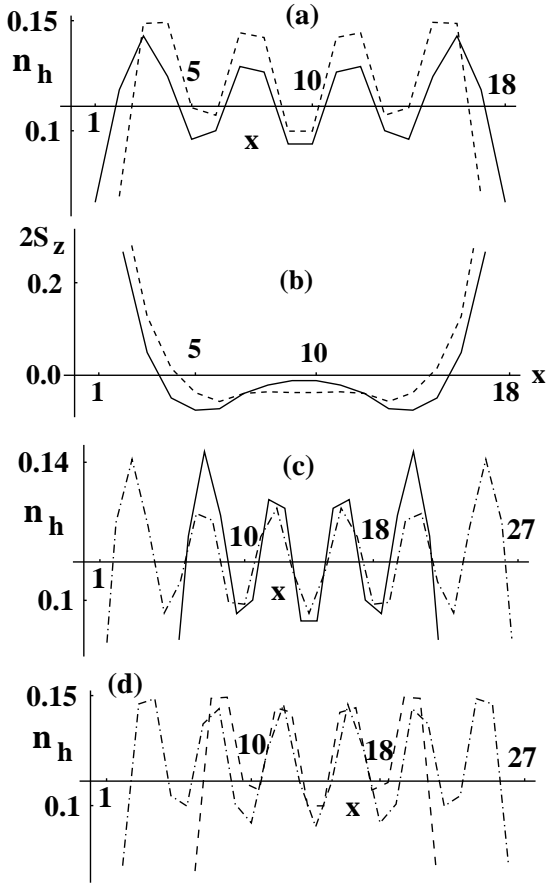}
\vspace*{.3cm}
\caption[]{ 
 (a) Hole density $\rho(x)$ and  (b) staggered spin
  density $S_z(x)$ as a function of
position $x$ along a 
$18 \times 4$ ladder ({\it i.e.}, with open boundary conditions in the $y$
direction)\cite{lrcoul}
with 8 holes ($n_h=1/9$).
The results with nonvanishing Coulomb prefactor ($V_0=t$,
solid lines) are compared with those without Coulomb interaction
 ($V_0=0$, dashed lines).
A comparison of
 $\rho(x)$ in a longer system ($12$ holes in $27
\times 4$ ) with the same $n_h=1/9$  (dashed-dotted line in both cases)
is shown in (c) and (d) respectively with and
without Coulomb interactions.
}
\label{18}
\end{figure}

\begin{figure}[htb]
\psfig{width=0.4\textwidth,file=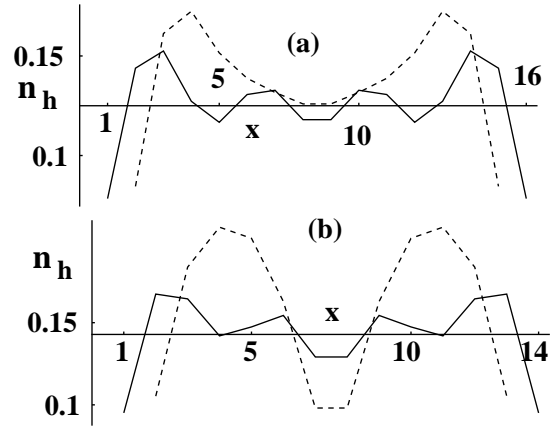}
\vspace*{.3cm}
\caption[]{ 
 Hole density $\rho(x)$ for 8 holes in a $16\times4$ (a) and a
$14\times4$ (b) ladder. Conventions are as in Fig.~\ref{18}.
\label{16}
}
\end{figure}

\begin{figure}[htb]
\psfig{width=0.4\textwidth,file=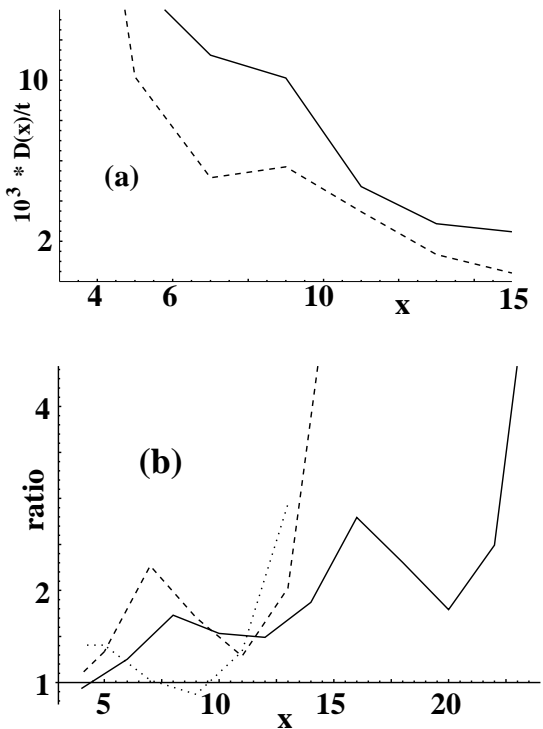}
\vspace*{.5cm}
\caption[]{
(a) The long-distance part  of the 
pair-field correlation $D(x)$ for 8 holes in a  $18
  \times 4$  ladder
without ${V}_{\mathrm Coul}$ (dashed line) and with ${V}_{\mathrm
  Coul}$ 
(solid line). 
 (b)  The ratio between $D(x)$ with ${V}_{\mathrm Coul}$
and without  ${V}_{\mathrm Coul}$ 
 for $N \times 4$ ladders with 8 holes and $N=18$ (dashed)
$N=16$ (dotted), and $N=27$ with 12 holes (solid){\protect{\cite{pair}}}.
}
\label{dl}
\end{figure}

Consider first the results of the DMRG calculations
without the long-range Coulomb potential. Results for the spin and charge
density are shown as the dashed lines  in
Figs.~\ref{18} and \ref{16} 
and for the pairing
susceptibility 
$D(x)$  by the  
dashed line in Fig.~\ref{dl}a.  

In the low-doping case,
  ($n_h=1/9$), depicted in Fig.~\ref{18}, 
charge stripe order
is clearly seen.
Stripes form in the $t$-$J$
model so as to satisfy the competing requirements of minimizing the kinetic
energy of the doped holes and minimizing the disturbance of the background exchange
interactions~\cite{wh.sc,zaanen}. 
In order to distinguish between stripes and ordinary Friedel oscillations,
we have compared the hole-density profile
for different length ladders  ($18 \times 4$ and $27 \times 4$)
at the same doping. As one can see from
Figs.~\ref{18}c and Figs.~\ref{18}d for the cases with and without
Coulomb interaction, respectively, the amplitude in the center of the
system is essentially independent of the system size,
while Friedel oscillations should decay as a function of the distance
from the boundary~\cite{Hell_vs_White,friedel}.
As seen in Fig.~\ref{18}b, any
$\pi$ phase shift in the AF order (spin stripes) is quite weak in the present
calculation, although it is slightly enhanced by the Coulomb interaction.
In accordance with Refs.~\cite{wh.sc} for the bare $t-J$
model,
spin stripe order is 
stronger in the case of cylindrical boundary conditions.
There 
is roughly one hole per two stripe unit cells,
which
was taken by WS\cite{wh.sc} as evidence that 
the $t-J$ model favors stripes with 
a minimum energy for a linear charge density of $\lambda \sim 0.5$. This is in
agreement with experiments {\cite{tr.st.95,tranquada,mook2000}},
 which find stripes with
$\lambda\approx 0.5$ at hole dopings smaller than 
$n_{h,c} \approx 1/8$~\cite{th.bi.89,ch.ae.91,za.ed.00,see_also}.

For $n_h \ge 1/8$ the density-wave structure seen in Fig.~\ref{16} 
 is less clear.  
Were the stripes to retain their integrity at these higher hole concentrations,
they would be forced very closer together, at considerable cost in energy.
 However, the clustering tendency of holes is still
apparent in the ground-state charge distribution, which is suggestive of two
hole-rich puddles, each with four holes. 

As discussed by WS, 
the hole clusters locally
share a number of features with the two-hole pair state, which accounts
for the fact that the energy per hole for a domain wall is close to the energy per
hole for a pair~\cite{wh.sc}. This is suggestive of a competition between 
stripe stability and superconductivity\cite{emery97}. Such a competition has
already been demonstrated in a model which includes next-nearest-neighbor
hoppings
$t'$ by WS.
For large enough $|t'|$,
the domain walls ``evaporate'' into quasiparticles ($t'<0$) without
significant pairing correlations or into pairs ($t'>0$)~\cite{wh.sc}.
 It has long been clear that stripe formation suppresses long-range superconducting
phase coherence\cite{kivelson98}, as is clear from the rapid falloff with distance of the
pair-field correlator in Fig.~\ref{dl}a (dashed line).

We now turn to our results in the presence of the Coulomb interactions.
Unscreened Coulomb interactions ($\varepsilon=1$, i. e.
 $V_0\simeq8.5t$) are so strong that they entirely dominate the physics,
destroying all clustering or pairing tendencies of the system.
However, for the physically relevant case, $\varepsilon\sim 8.5$,
the results are much more interesting.

In the lightly doped case, $n_h=1/9$,
 shown in Fig.~\ref{18}, the stripe structure is essentially
unchanged by Coulomb interactions, although  the amplitude of the charge
modulations is suppressed (by roughly a factor of $1.5$), and the anti-phase
character of the spin correlations is slightly enhanced.  We interpret 
this as meaning that the stripe order is robust, 
but that the Coulomb interactions
enhance the transverse stripe fluctuations.  

The most dramatic effect of the Coulomb interactions is 
the strong enhancement of the pair-field correlations shown in
Fig.~\ref{dl}.
From Fig.\ref{dl} (in which the short-distance data is off scale)  one might 
conclude
 that the Coulomb interactions primarily increase the
{\it overall}
 magnitude of the pair correlations.
On the other hand, 
 from the ratio between the two functions, which is displayed 
in Fig.~\ref{dl}b over the whole range of $x$,
 one can see that it is only the {\it
   long-distance}
 part that is enhanced. 
In order to sort out boundary effects,
which are probably responsible for part of the strong increase of the
ratio at the longest distances available,
 we have also treated a
larger ($27\times4$) ladder.  Although
the calculation of $D(x)$ is less accurate for this
system~\cite{pair}, 
we can still 
draw some conclusions.
As one can see from the figure,
the ratio of $D(x)$ shows oscillations with twice the stripe periodicity, 
whose envelope is clearly increasing with distance, even far away from
the boundaries.
 The dependence of $D(x)$ on distance, seen in Fig.~\ref{dl}a, is altered
from rapidly decreasing in the absence of Coulomb interactions, to much slower
distance dependence in the presence of Coulomb interactions.  
 This result
supports the idea that while 
 the longer-range phase coherence (or, in
other words, pair delocalization) is inhibited by rigid stripe order,
as obtained in the pure $t-J$ model,   
{\it stripe
fluctuations, induced by the Coulomb interaction, permit the pairs to
tunnel from stripe to 
stripe.
}   
It should be pointed out that this conclusion may not be generic and
may depend on the stripes stiffness.
If stripes are intrinsically weak, we expect their fluctuations to be strong {\it
  without} the need of Coulomb interaction.
In this case, the pair-breaking effects of the Coulomb interaction
would probably obtain the opposite effect and suppress pairing correlations.
It could also be the case that in our relatively narrow systems
stripes are particularly stiff, due to the fact that they cannot
meander effectively.

To further corroborate this interpretation, we have 
computed the (finite-size) 
spin gap with and without Coulomb interactions  for
the $N=14$ system (with $h=0$). 
This  is $\Delta_{s}\sim 0.16 t$ without Coulomb 
interaction, and $\Delta_{s}\sim 0.12 t$ with $V_0=t$. 
Since we know from the work of WS~\cite{wh.sc}, that the local stripe
energetics is essentially dominated by the short-range pair binding (which
sets a scale for the spin gap as an $S=1$ excitation of the pair) and
since we find $\Delta_s$ only slightly reduced
 in the presence of
 Coulomb interaction, it is reasonable to assume that the local
 pairing itself is still due to the short-range $t-J$ physics.

The effects of Coulomb interactions on the charge-density profile 
are even stronger
(and more complex) in the more heavily doped  
systems as shown for $N=16$  
($n_h=1/8$)  and $N=14$ ($n_h=1/7$) in Fig.~\ref{16}.
In these systems,
even the period of the stripe array is altered by the Coulomb
interactions. In particular, the two hole puddles  seen as 
minima in the electronic density in the absence 
of Coulomb interactions are broken apart into structures that look 
somewhat more like the four-stripe pattern seen in the $N=18$ ladder.
This result is an example of a period selection caused by
a Coulomb-frustrated tendency to clustering, or phase 
separation\cite{lo.em.94}.
In this case, we expect particularly large fluctuations between
the competing configurations to enhance the delocalization of pairs 
between stripes.  

\section{Conclusions}
\label{conc}
In summary, in this paper we present a new approach capable of
bridging the gap between work~\cite{lo.em.94,emery97} arguing that
stripes are due to a delicate balance between kinetic and long-range
Coulomb interaction energy, and work~\cite{wh.sc} support the idea
that 
short-range interactions alone can lead to stripe formation. Based on
a new technique (``density-functional'' DMRG), we present numerical
results, supporting the view that---while short-range
``$t$-$J$-like'' interactions locally bind holes into pairs---it is
the long-range Coulomb interaction which induces their delocalization
accompanied with substantial enhancement of superconducting pairing
correlations.  Moreover, the Coulomb-induced stripe fluctuations 
suppress the magnitude of charge-density wave order,
but can actually slightly enhance the (``anti-phase'') spin-density wave correlations with
twice the wavelength.  

\section*{Acknowledgments}
We
 are grateful for fruitful discussions with W. Kohn and
D. J. Scalapino, as well as support with the DMRG code by R. M. Noack.
We acknowledge financial support from the DFG
(HA 1537/14-1, 
HA 1537/17-1, and HA 1537/20-1). 
SAK was supported in part by NSF grant number DMR98-08685  at UCLA.
WH and SAK are grateful to the ITP in Santa Barbara for generous
hospitality and financial support by NSF grant No. PHY99-07949.

\appendix

\section{Validity of the Hartree approximation}
\label{appa}
The 
Coulomb interaction part of the Hamiltonian reads
\beq
 W = \frac12 \sum_{{\ve r}\not= {\ve r}'} \frac{V_0}{|{\ve r}-{\ve r}'|} \  n_{\ve r} \  n_{{\ve r}'} \;,
\eeq
where $ n_{\ve r}$ is the density operator.
This can be rewritten as
\begin{eqnarray}
 W &&= \sum_{{\ve r}\not= {\ve r}'} \frac{V_0}{|{\ve r}-{\ve r}'|} \ <n_{\ve r}> \  n_{{\ve r}'} 
\\ \nonumber &&
\label{w}
+
\frac12 \sum_{{\ve r}\not= {\ve r}'} \frac{V_0}{|{\ve r}-{\ve r}'|} \ ( n_{\ve r} - <n_{\ve r}>)
\ ( n_{{\ve r}'} - <n_{{\ve r}'}>) 
\\ \nonumber &&
+ \hbox{constant} \;,
\end{eqnarray}
where $<n_{\ve r}>$ is the ground-state expectation value of $ n_{\ve r}$.
Here, the first term on the r.h.s. of \eqref{w} (let's us call it $W_H$)
corresponds to the Hartree approximation, while the second ($W_c$) gives
its correction.
Of course, when the expectation value of $W_c$ vanishes, the
Hartree approximation becomes exact (for ground-state properties). 
Since, by definition, $g({\ve r},{\ve r}') \equiv \frac{< n_{\ve r}  n_{{\ve r}'}>}{<n_{\ve r}>\
  <n_{{\ve r}'}>}$,
$<W_c>$ can be also written as
\beq
<W_c> = \frac12 \sum_{{\ve r}\not= {\ve r}'} \frac{V_0}{|{\ve r}-{\ve r}'|} \ <n_{\ve r}>\  <n_{{\ve r}'}> ( g({\ve r},{\ve r}') -1 )\;,
\label{wc}
\eeq
and, thus, the correction to Hartree vanishes when $g({\ve r},{\ve r}')=1$.

In order to  measure the accuracy of a mean-field approximation 
for a generic operator $\hat O({\ve r}) \ \hat O({\ve r}')$ in the Hamiltonian, 
one should evaluate its fluctuations
\beq
\Delta_O({\ve r},{\ve r}') \equiv \langle (\hat O({\ve r})-<\hat O({\ve r})>) \ (\hat
O({\ve r}')-<\hat O({\ve r}')>)\rangle \;.
\eeq
Since these terms gives the corrections to the mean-field
approximation (cf. \eqref{wc}), a small value for $\Delta_O$ means
that the mean-field approximation for $O$ is accurate.

In our calculations, 
 the {\bf density} fluctuations $\Delta_n({\ve r},{\ve r}')$ are typically about $20-30$
times smaller than the {\bf spin } fluctuations $\Delta_{\ve
  S}({\ve r},{\ve r}')$,
for nearest-neighbor ${\ve r}$ and ${\ve r}'$, where
both fluctuations are largest (for ${\ve r}\not={\ve r}'$). 
{ This is
the reason for using the Hartree approximation for the Coulomb
part and to treat the ``$J$'' part exactly}.
Similarly, on the same site, $\Delta_n({\ve r},{\ve r})\approx 0.1$, which
is of the same order as $\Delta_{\ve S}({\ve r},{\ve r}')$, which, in turn, justifies
an exact treatment of the {\it on-site} interaction.

Notice that the inhomogeneous state brought about by the open boundary
conditions is decisive in making the Hartree approximation work
better.
Indeed, in a homogeneous system (obtained by periodic b.c.) $<n_{\ve r}>$
would be constant, and the oscillations in the charge will be all
shifted to the fluctuations $ n_{\ve r}-<n_{\ve r}>$ , making the correction $\Delta_n({\ve r},{\ve r}')$
large. On the other hand, for open boundary conditions, part of the charge
oscillations are taken care of by the mean values $<n_{\ve r}>$ making 
$ n_{\ve r}-<n_{\ve r}>$ (and thus $\Delta_n({\ve r},{\ve r}')$) smaller.





\bibliographystyle{prsty}
\bibliography{footnotes}

\begin{thebibliography}{10}
\bibitem{lrcoul}
While the short-range part of the Hamiltonian has open or periodic boundary
  conditions, the long-range Coulomb potential \eqref{coul} is always evaluated
  by repeating the cluster periodically in the 2D plane.
\bibitem{newton}
In practice, the self-consistent procedure has been speeded up by using a
  variant of the Newton method.
\end{thebibliography}

\end{document}